\documentclass[epsf,graphics,a4,12pt]{article}
\usepackage{epsfig}
\begin{document}
\baselineskip=18pt
\newcommand{\be}{\begin{equation}}
\newcommand{\en}{\end{equation}}
\begin{center}
{\large Holography in an Early Universe with Asymmetric Inflation}
\end{center}
\vspace{1ex}
\centerline{\large
Elcio Abdalla and L. Alejandro Correa-Borbonet}
\begin{center}
{Instituto de F\'{\i}sica, Universidade de S\~{a}o Paulo,\\
C.P.66.318, CEP 05315-970, S\~{a}o Paulo, Brazil}
\end{center}
\vspace{6ex}
\begin{abstract}
We discuss the validity of the holographic principle in a $(4+n)$
dimensional universe in  an asymmetric inflationary phase.
\end{abstract}
\vspace{2ex} \hspace*{0mm} PACS number(s): 98.80.Cq 04.50.+h
\vspace{2ex} 
%\vfill
%\newpage

Recently a new possible solution to the hierarchy problem has been
proposed \cite{Arkan}, where the fundamental Planck scale goes down to the
level of the Tev region, provided that there are compact extra dimensions 
where gravitationals fields propagate. In such a case, the size of these
extra dimensions should
be in the submillimiter scale in order to conform to the phenomenological
constraints.  Indeed, using Gauss's law in $(4+n)$
dimensions it is possible to relate the Planck scale in the $(4+n)$
dimensional theory($M_{*}$) with that of the $4$-dimensional
theory ($M_{Pl}$) by means of the size of the $n$ compactified dimensions
($b_{0}$). It is easy to find \cite{Dimo}
\be
M^{2}_{Pl}=(b_{0})^{n}M^{n+2}_{*} %%%%% (1)
\en
Assuming the usual value $M_{Pl}\simeq 10^{19}$ GeV and $M_{*}\simeq 1$
TeV, we find $b_{0}\sim 10^{-17+\frac{30}{n}}$ cm. For $n=2$ we obtain the
(reasonable) value $b_{0}\simeq 1$ mm. We would thus find deviation 
from Newtonian gravitation in a range within experimental search in the 
near future. 

There are several papers discussing the theoretical and phenomenological
aspects of this idea \cite{vario}. From the fundamental point of view, 
gravity can propagate freely around the whole set of dimensions in the
submillimeter region, but the Standard Model (SM) fields get confined in the
$4$-dimensional world. This is feasible if we consider that the SM fields
are localized on a $3$-brane immersed in the higher-dimensional space, 
as in the case of D-branes occurring in type I/II string  
theory\cite{Polch},\cite{Anton}. Thus a very attractive model of early
universe has been proposed \cite{Dimo}, and 
the underlying cosmology that occurs for small extra dimensions can be of
wide interest.

In this model the dynamics of the dimensions as described by the Einstein
equations is the central question. They are divided in a (so called
wall) four dimensional world, and the $n$-internal dimensions. In concrete
terms, the picture proposed is that of an early inflationary era followed
by a long epoch where the scale factor of the brane-universe undergoes a
slow contraction while the internal dimensions continue to expand towards 
their stabilized value. When the contraction ends the universe goes
asymptotically to the radiation dominated
be\-ha\-viour as described by the FRW model. Although details of the
dynamics underlying the transition are 
not known, a reasonable alternative explanation of the
dynamics leading to the Cosmological standard model may arises.

On the other hand, we have witnessed a great activity on the holographic
principle \cite{hooft,sussk}
which states that all information about physical processes in cosmology can be
stored on the surface enclosing the physical universe. This idea has
received strong  support from another conjeture known as the AdS/CFT
correspondence\cite{Malda}. The conjetured equivalence between Supergravity 
on D-di\-men\-sio\-nal AdS space and conformal field theory on the 
$(D-1)$-dimensional boundary has been proven very useful to gain
information about supergravity in the bulk from knowledge about its
boundary. Moreover, the holographic principle has been extended to
cosmology and  there have been many attempts to apply different
formulations of this principle to various cosmological models.

In this paper we will try to check the compatibility of the holographic
principle with the cosmological model proposed by Arkani-Hamed et
al \cite{Dimo}. Specifically the cosmological version of the holographic
principle proposed by Fischler and Susskind\cite{hol} will be checked. 
Here we will restrict to the study of the contraction epoch
with two extra dimensions, in agreement with the afore mentioned
phenomenological constraints.

The evolution of the scale factors $a(t)$ corresponding to the
$3$-dimensional physical 
(observable) space, and $b(t)$, corresponding to the internal dimensions, is
described by the Einstein equations as given by\cite{Dimo}
\be
6H^{2}_{a}+2H^{2}_{b}+12H_{a}H_{b}=\frac{V}{M^{4}_{*}b^{2}} , %%%%% (2)
\en
\be
\frac{\ddot{b}}{b}+H^{2}_{b}+3H_{b}H_{a}=
\frac{1}{M^{4}_{*}b^{2}}(\frac{V}{2}-\frac{b}{8}\frac{\partial V}{\partial
  b})  , %%%%% (3)
\en
\be
\frac{\ddot{a}}{a}+2H^{2}_{a}+2H_{b}H_{a}=
\frac{b}{8M^{4}_{*}b^{2}}\frac{\partial  V}{\partial b} .  %%%%% (4)
\en
where we have introduced the Hubble parameters $H_{a}\equiv \dot{a}/a$ and
$H_{b} 
\equiv \dot{b}/b$ for the two scale factors, the overdot denoting the
derivative with respect to $t$. Here, $V$ is the stabilizing potential,
which is a function of the scale factor $b$. These quantities are in the
socalled string (brane) frame. The 
reason is that the kinematics in this frame is automatically expressed in
terms of the units measured by the observers which live on the wall.

This system has been obtained from the original Einstein equations after
neglecting the energy density $\rho$ and pressure $p$ (see Eq($57$) in
\cite{Dimo}).
Using the COBE data we can prove that values of $\rho$ and $p$ are much
smaller than the
potential $V$. This potential, in the semiclassical limit, may be viewed as
an expansion in inverse powers of the scale factor $b(t)$. Simplifying, it
has been taken as being a monomial form, $V=Wb^{-p}$, where $W$ is a
dimensionful parameter.

These equations can be solved exactly. Defining
\be
a=a_{i}e^{\alpha (t)}   \;\;\;\;\;\;   b=b_{i}e^{\beta (t)}  %%%%% (5)
\en
and going to the new time variable
\be
d\tau=-e^{-3\alpha-2\beta}dt %%%%% (6)
\en
the previous system, after some algebra, can be reduced to a functional
constraint equation \cite{Dimo}
\be
4\alpha+\frac{4p}{4+p}\beta=C_{1}+C_{2}\tau %%%%% (7)
\en
together with a simple first order differential equation,
\be
X'^{2}=\frac{\Delta}{4}\omega e^{X}+\frac{3(4+p)^{2}}{32}C^{2}_{2} %%%%% (8)
\en
where
\be
X=6\alpha +(2-p)\beta %%%%% (9)
\en
and $C_{1}$, $C_{2}$ are integration constants. The other two parameters that
appear in the system are $\Delta=8-8p-p^{2}$ ,
$\omega =\frac{W}{M^{4}_{*}b^{2+p}_{i}}$. The prime denote the derivative
with respect to $\tau$.

The solutions of the system take different form, controlled by whether
$C_{2}$ vanishes
or not, and by the value of $\Delta$. It is found that the solutions can
be classified in four
types: 
\be
(1)\,\Delta=0; \quad   (2)\,\Delta>0,C_{2}=0; \quad (3)\,\Delta>0,C_{2}<0 ;
\quad (4)\,\Delta<0,C_{2}<0 %%%%% (10)
\en
The critical value $\Delta=0$ corresponds to $p\simeq0.899$. In this
case the solution is
\be
a=a_{i}(\frac{t}{t_{i}})^{k}\quad ,\label{eq:facta} %%%%% (11)
\en
\be
b=b_{i}(\frac{t}{t_{i}})^{l}\quad ,\label{eq:factob} %%%%% (12)
\en
with $k=-0.1266$ and $l=0.69$. Such values satisfy the two algebraic
equations, namely
\be
3k+2l=1\label{eq:kasner}\quad ,
\en
and
\be
3k^{2}+2l^{2}=1\quad ,
\en
implying that in this era the solution describes a Kasner universe.

Applying the F-S criteria for anisotropic universes the following 
entropy/area relation is obtained, with the {\it bar} quantities defined
in the Einstein frame,
\be
S/A=\frac{\bar{\sigma}  (\bar{R}_{H,\bar{a}})^{3}(\bar{R}_{H,\bar{b}})^{2}}
{[(\bar{a}\bar{R}_{H,\bar{a}})^{3}(\bar{b}\bar{R}_{H,\bar{b}})^{2}]^{4/5}}
\label{eq:einsfram}\quad ,
\en
where the comoving size of the horizon is
\be
\bar{R}_{H,\bar{a}}(\bar{t})=\int^{\bar{t}}_{\bar{t}_{i}} 
\frac{d\bar{t}'}{\bar{a}(\bar{t}')}\label{horizon},
\en
and $\bar{\sigma}$ is the comoving entropy density.

In order to calculate the S/A relation $(\ref{eq:einsfram})$ we use the
expression that relate the metric in the Einstein frame with the one in
the string frame \cite{Dimo} 
\be
\bar{g}_{\mu \nu}=b^{2}g_{\mu \nu} \quad .
\en
Using this map, we get
\be
\bar{a}=ba   \;\;\;\;\;\;   \bar{b}=bb \;\;\;\;\;\; d\bar{t}=bdt \quad .
\en
Therefore we have
\be
\bar{R}_{H,\bar{a}}=R_{H,a} \;\;\;\;\;\;\;  \bar{R}_{H,\bar{b}}=R_{H,b} \quad .
\en
Now, the S/A relation can be rewritten as 
\be
S/A=\frac{1}{b^{4}}\frac{\sigma  (R_{H,a})^{3}(R_{H,b})^{2}}
{[(aR_{H,a})^{3}(bR_{H,b})^{2}]^{4/5}}
\label{eq:stringfram} \quad .
\en
The authors of Ref \cite{Kalin} argued that when the Fischler-Susskind
criteria is
applied to inflationary cosmology the evolution of the universe should be
considered
after reheating. For the model of Arkani-Hamed et $al$ we will take the
initial time
$t_{i}$ as corresponding to the end of the de Sitter phase, because the
model does not have
the usual post-inflationary reheating of standard inflationary
cosmology. We also
should remember that the end of the de Sitter phase is the beginning of
the contraction epoch.

Taking into account the scale factors $(\ref{eq:facta})$ and 
$(\ref{eq:factob})$ we get, after some manipulations,
\be
S/A=\frac{\sigma}{a^{3}_{i}b^{6}_{i}}c_{1}t_{i}f(x),
\en
where
\be
c_{1}=\frac{1}{(1-k)^{3/5}}\frac{1}{(1-l)^{2/5}}=1,4873
\en
and
\be
f(x)=\frac{1}{x^{4l}}\frac{(x^{1-k}-1)^{3}(x^{1-l}-1)
^{2}}{(x-x^{k})^{12/5}(x-x^{l})^{8/5}} \quad , 
\en
where $x=t/t_{i}$.

The shape of $f(x)$ is illustrated in figure 1. Using $(\ref{eq:kasner})$
it can be checked that the asymptotic value of $f(x)$ is zero. Therefore if 
the holographic bound holds when $f(x)$ has its maximum value, later on, the 
bound will be satisfied even better. 

The solutions in the case $\Delta>0$($0\leq p<0.899$) and $C_{2}=0$ are
similar but the corresponding powers are
\be
v=-\frac{2p}{(2+p)(4+p)}\label{eq:expv} \quad ,
\en
\be
u=\frac{2}{2+p}\quad .\label{eq:expu}
\en
In this case the powers satisfy
\be
3v+2u>1\quad .
\en
Repeating the same procedure as the previous case we find for the entropy
to area ratio the result
\be
S/A=\frac{\sigma}{a^{3}_{i}b^{6}_{i}}c_{2}t_{i}g(x)
\label{eq:entrorA}\quad ,
\en
where for $p<0.899$
\be
c_{2}=\frac{1}{(1-v)^{3/5}}\frac{1}{(1-u)^{2/5}}\quad ,
\en
\be
g(x)=\frac{1}{x^{4u}}\frac{(x^{1-v}-1)^{3}(x^{1-u}-1)^{2}}{(x-x^{v})
^{12/5}(x-x^{u})^{8/5}}\quad ,
\en
and for $p=0$\quad ,
\be
c_{2}=1
\en
\be
g(x)=\frac{1}{x^{4}}\frac{(x-1)^{3/5}(ln(x))^{2/5}}{x^{8/5}}\quad .
\en

In figure 2 the function $g(x)$ is shown for diferents values of the
parameter $p$($0,0.5,0.7$). Also in this case the asymptotic value of $g(x)$ 
is zero. In figure 3 a comparation between the functions $f(x)$ and $g(x)$ is 
displayed.

In the case ($\Delta<0 (p>0.899), C_{2}<0 $)the Kasner solution applies
asymptotically
(see Ref $\cite{Dimo}$). In other words, the scale factors $a(t)$ and $b(t)$
take the form $(\ref{eq:facta}),(\ref{eq:factob})$ for large times.
This implies that the ratio entropy/area will coincide with the $\Delta=0$ case
in the remote future (see figure 4).

The last solution ($\Delta>0,C_{2}<0$) is similarly reduced to the case
($\Delta>0, C_{2}=0$). Then, the scale factors will have the powers
$(\ref{eq:expv}),(\ref{eq:expu})$ and the entropy/area
relation will be described by equation $(\ref{eq:entrorA})$ in the remote
future.
\begin{figure}
\begin{center}
\leavevmode
\epsfig{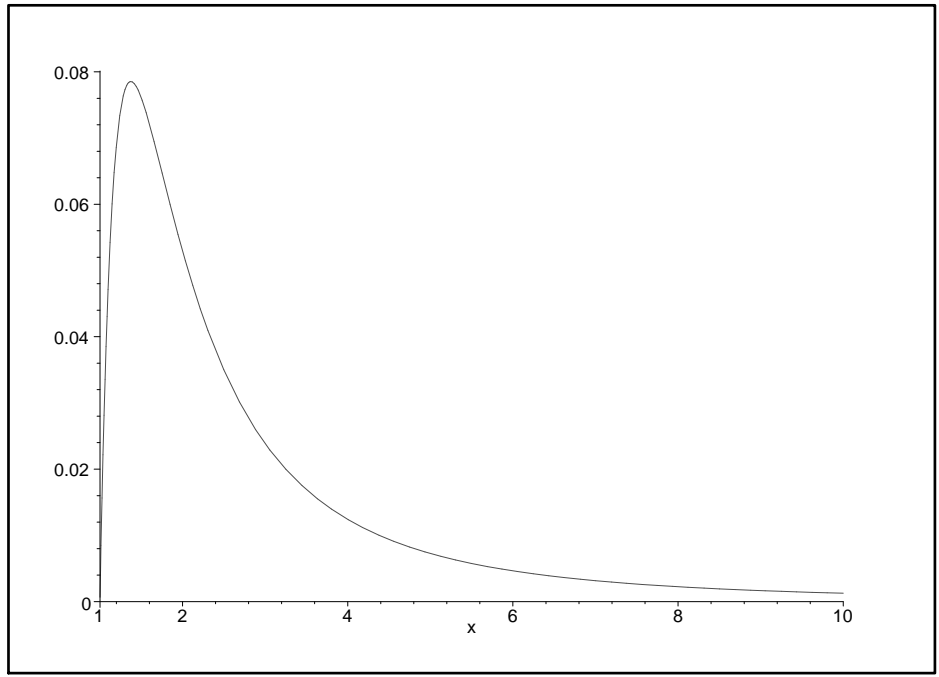}
\end{center}
\caption{Function f(x) in the critical case $p=0.899.$}
\label{a}
\end{figure}

%\begin{eqnarray}
%\epsfxsize= 8truecm\rotatebox{-90}{\epsfbox{primero101.eps}}\nonumber
%\end{eqnarray}
%\end{center}
%\vskip -1.5cm
%%\end{center}
%\caption{Function f(x) in the critical case, $p=0.899.$}
%\label{a}
%\end{figure}

\begin{figure}
\begin{center}
\leavevmode
\epsfig{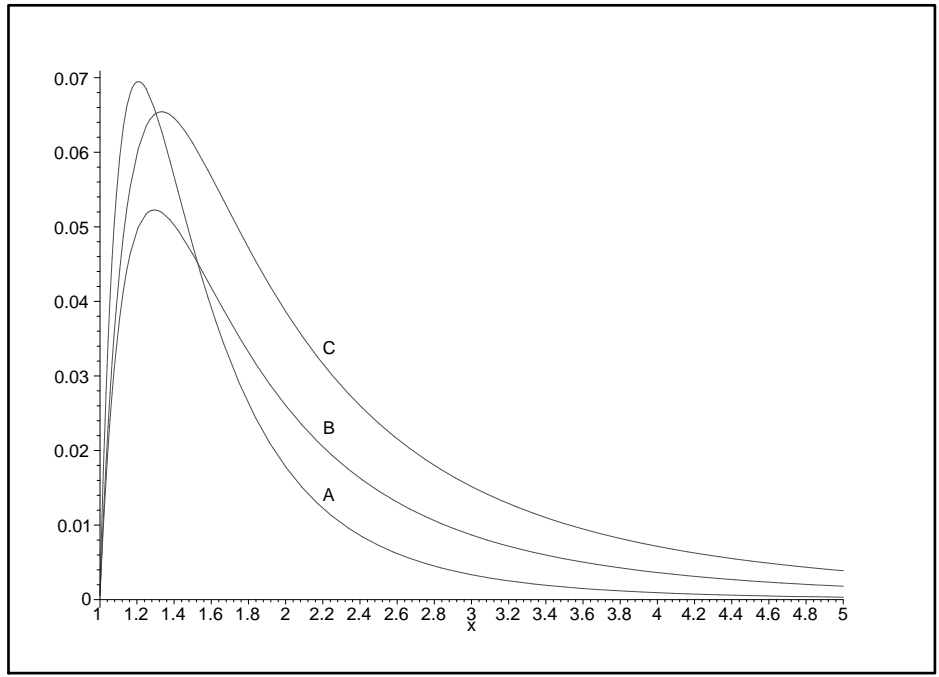}
\end{center}
\caption{Function g(x) in the subcritical cases $p=0(A),0.5(B),0.7(C)$.}
\label{b}
\end{figure}
\begin{figure}
\begin{center}
\leavevmode
\epsfig{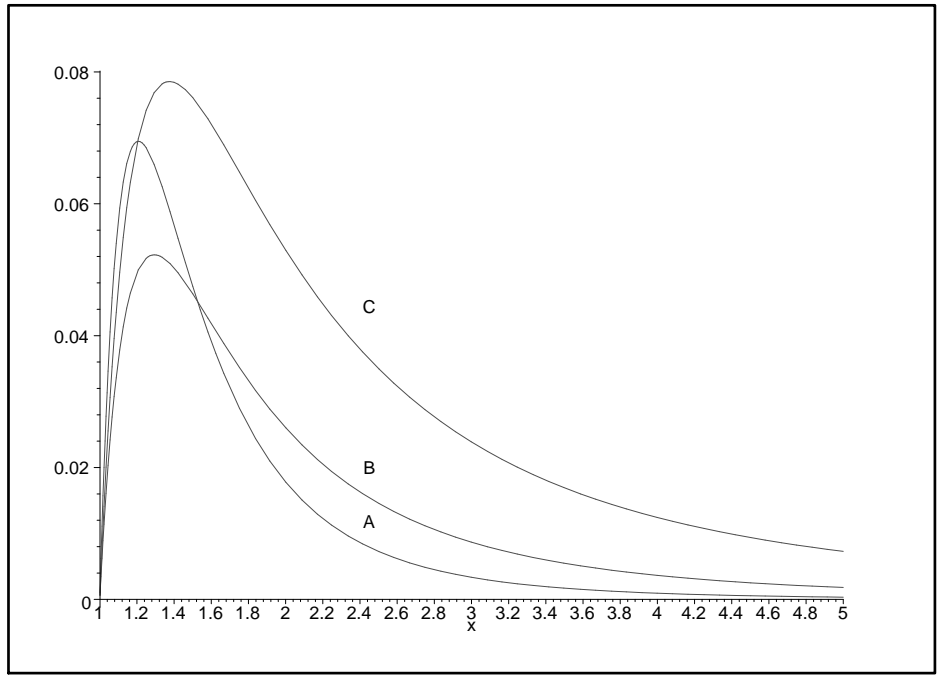}
\end{center}
%\end{center}
\caption{Comparison between the cases $p=0(A),0.5(B),0.899(C)$.}
\label{c}
\end{figure}
\begin{figure}
\begin{center}
%\mbox{\epsfig{file=graph301.eps,width=0.5\textwidth}}
\leavevmode
\epsfig{file=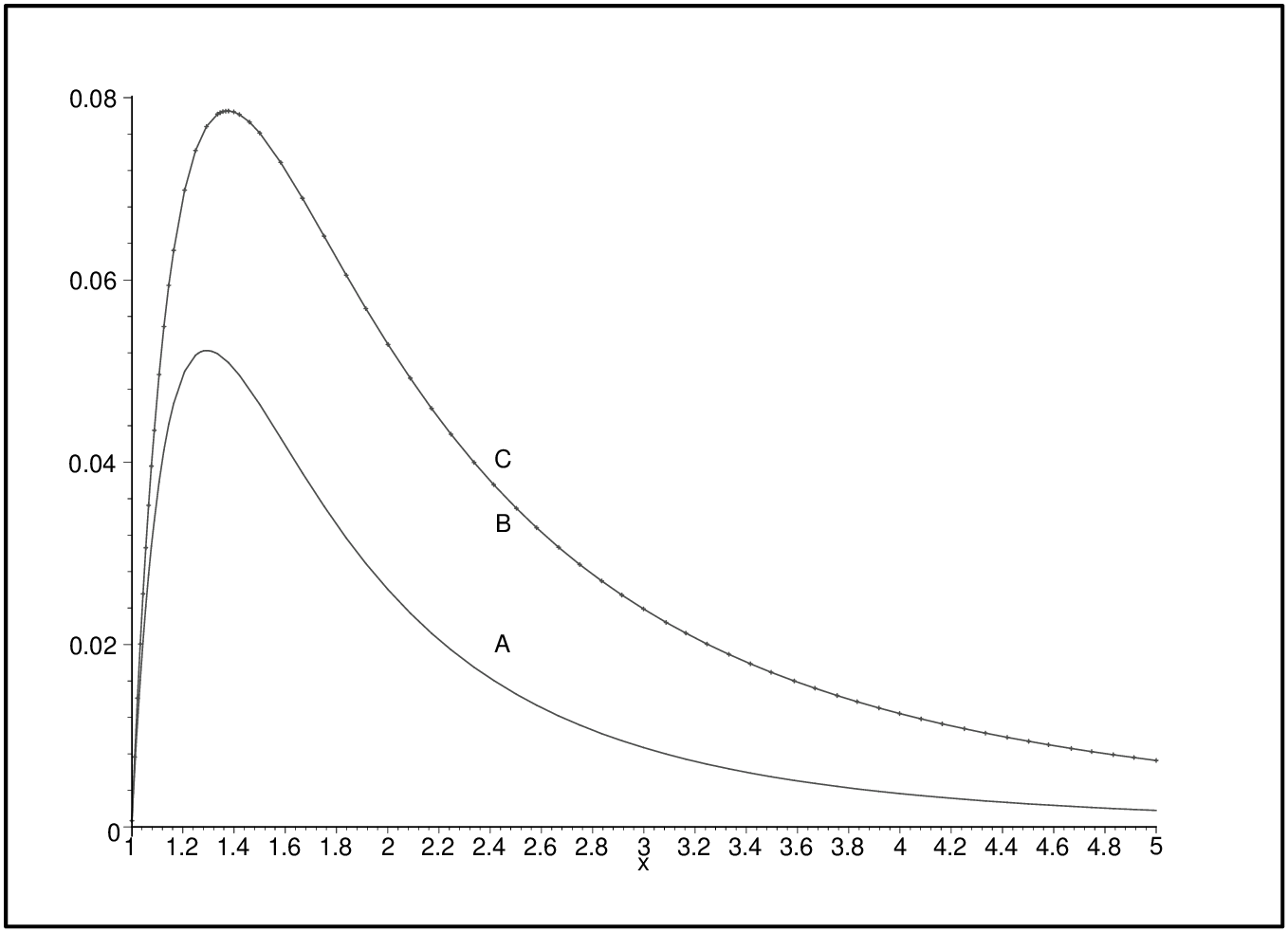,width=0.90\textwidth,angle=-90}
\end{center}
\vskip -1.5cm
%\end{center}
\caption{Comparison between the cases $p=0.5(A),0.899(B),8(C)$.}
\label{d}
\end{figure}

In conclusion, we have found that the holographic principle can be satisfied
in the contraction epoch for the model of Arkani Hamed et
$al$\cite{Dimo}. The entropy to area ratio has  a similar
asymptotic behavior for all the solutions. As this
model presents some novelties for inflationary cosmology, the results that
we have obtained could be useful in order to improve our understanding
of the holographic principle in relation to cosmology.
%\newpage

ACKNOWLEDGMENT:
This work was partially supported
by {\it Funda\c{c}\~ao de Amparo \`a Pesquisa do Estado de
S\~ao Paulo} (FAPESP), {\it Conselho Nacional de Desenvolvimento  
Cient\'{\i}fico e Tecnol\'{o}gico} (CNPQ) and CAPES (Ministry of Education,
Brazil).

%\newpage


\begin{thebibliography}{99}
\bibitem{Arkan} N. Arkani-Hamed, S. Dimopoulos and G.Dvali,
{\it Phys. Lett.} {\bf B429}, 263 (1998).
\bibitem{vario} P. Argyres, S. Dimopoulos and J. March-Russell, 
{\it Phys. Lett.} {\bf B441}, 96 (1998);
Z. Kakushadze and S.H. Tye, {\it Nucl. Phys.} {\bf B548}, 180 (1999);
K. Benakli, {\it Phy. Rev.} {\bf D60}, 104002 (1999); L.Randall and R. 
Sundrum, {\it Nucl.Phys.} {\bf B557}, 79 (1999).     
\bibitem{Polch} J. Polchinski, {\it Phys. Rev. Lett.} 75, {\bf 4724} (1995).
\bibitem{Anton} I. Antoniadis, N. Arkani-Hamed, S. Dimopoulos and G.Dvali,
{\it Phys. Lett.} {\bf B436}, 257 (1998).
\bibitem{Dimo} N. Arkani-Hamed, S. Dimopoulos, N. Kaloper and
J. March-Russell, {\it Nucl. Phys.} {\bf B567}, 189 (2000).
\bibitem{hooft} G.'t Hooft, gr-qc/9310026.
\bibitem{sussk} L.Susskind, {\it J. Math. Phys.} {\bf 36}, 6377 (1995).
\bibitem{Malda} J.M. Maldacena, {\it Adv. Theor. Math. Phys.} {\bf 2}, 231
(1998);   E.Witten, {\it Adv. Theor. Math. Phys.} {\bf 2}, 253 (1998); 
S.S.Gubser, I.R. Klebanov and A. M. Polyakov, {\it Phys. Lett.} {\bf
B428}, 105 (1998). 
\bibitem{hol} W.Fischler and L. Susskind, Holography and Cosmology,
hep-th/9806039. 
\bibitem{Kalin}N.Kaloper and A.Linde, {\it Phys.Rev.} {\bf D60}, 103509 (1999).
\end{thebibliography}
\end{document}